\documentclass[aps,amsfonts,nofootinbib,preprintnumbers,superscriptaddress,nobalancelastpage]
{revtex4}
\pdfoutput=1
\usepackage{graphicx}
\usepackage[fleqn]{amsmath}
\usepackage{amssymb}
\usepackage{url}
\usepackage{array,multirow}
\usepackage{hyperref,color}
\usepackage[normalem]{ulem}
\usepackage{slashed}
\usepackage{rotating}
\usepackage{afterpage}
\usepackage{ifthen}
\usepackage{mciteplus}
\usepackage{color}
\usepackage{adjustbox}

\usepackage[switch]{lineno}

\newcommand{\sla}[1]{/\!\!\!#1}

\def\lsim{\raise0.3ex\hbox{$\;<$\kern-0.75em\raise-1.1ex\hbox{$\sim\;$}}}
\def\gsim{\raise0.3ex\hbox{$\;>$\kern-0.75em\raise-1.1ex\hbox{$\sim\;$}}}

\newcommand {\owww}    {{\cal O}_{WWW}}
\newcommand {\ow}      {{\cal O}_{W}}
\newcommand {\ob}      {{\cal O}_{B}}
\newcommand{\eh}{\hat{e}}
\newcommand{\sh}{\hat{s}}
\newcommand{\ch}{\hat{c}}
\newcommand{\vh}{\hat{v}}

\begin{document}
\preprint{HRI-RECAPP-2024-03} 

\title{Triple gauge coupling analysis using boosted $W$'s and $Z$'s}
\author{O.\ J.\ P.\ \'Eboli}
\email{eboli@if.usp.br}
\affiliation{Instituto de F\'{\i}sica, 
Universidade de S\~ao Paulo, S\~ao Paulo -- S\~ao Paulo 05580-090, Brazil.}
\author{Tathagata Ghosh}
\email{tathagataghosh@hri.res.in}
\affiliation{Regional Centre for Accelerator-based Particle Physics, Harish-Chandra Research Institute, A CI of Homi Bhabha National Institute,
  Chhatnag Road, Jhunsi, Allahabad, 211019, India}
\author{Matheus Martines}
\email{matheus.martines.silva@usp.br}
\affiliation{Instituto de F\'{\i}sica, 
Universidade de S\~ao Paulo, S\~ao Paulo -- S\~ao Paulo 05580-090, Brazil.}
\author{Sujay Shil}
\email{sujayshil1@gmail.com}
\affiliation{Instituto de F\'{\i}sica, 
Universidade de S\~ao Paulo, S\~ao Paulo -- S\~ao Paulo 05580-090, Brazil.}

\begin{abstract}  

  We analyze the Large Hadron Collider potential to study triple
  couplings of the electroweak gauge bosons using their boosted
  hadronic decays. Deviations from Standard Model predictions spoil
  cancelations present in the Standard Model leading to the growth of
  the electroweak diboson production cross section at high
  center-of-mass energies. In this kinematical limit, $W$'s and $Z$'s
  are highly boosted, and consequently, their hadronic decays give
  rise to fat jets. Here, we show that the study of boosted
  hadronically decaying $W$ and $Z$ leads to limits on triple gauge
  couplings that are comparable to the ones originating from the
  leptonic decay channels.

\end{abstract}

\maketitle

\section{Introduction}
\label{sec:intro}

The CERN Large Hadron Collider (LHC) has already accumulated a
substantial dataset allowing for precision tests of the Standard Model
(SM) and searches for new physics.  Within the SM framework, the
triple and quartic vector-boson couplings are determined by the
non-abelian $SU(2)_L \otimes U(1)_Y$ gauge symmetry, being completely
fixed in terms of the gauge couplings.  Possible deviations from the
SM predictions for the triple gauge couplings (TGC) are a clear sign
of new physics and consequently, TGCs are investigated at the LHC in
charged processes like $W^+W^-$, $W^\pm Z$ and $W^\pm \gamma$
productions. \smallskip


Anomalous TGC can be generated by integrating out new heavy states,
however, the resulting anomalous couplings are suppressed by loop
factors $1/(16\pi^2)$ for weakly interacting extensions of the
SM~\cite{Arzt:1994gp}. Notwithstanding, it is possible to write down
ultraviolet extensions of the SM that generate anomalous TGC at tree level and
that are not constrained by the electroweak precision
observables~\cite{Falkowski:2016cxu, Arkani-Hamed:2001kyx}. \smallskip

Due to the importance of TGC measurements, the ATLAS and CMS
collaborations conduct studies on the electroweak TGC through various
processes. They analyze the
$W^\pm Z \to \ell^+ \ell^- \ell^{\prime \pm} + E_T^{\text{miss}}$
channel~\cite{ATLAS:2018ogj, CMS:2021lix} as well as the
$W^+ W^- \to \ell^+ \ell^{(\prime) -} + E_T^{\text{miss}}$ final
state~\cite{CMS:2020mxy, ATLAS:2021jgw}. Additionally, TGC were
investigated in the
$W^\pm \gamma \to \ell^\pm \gamma + E_T^{\text{miss}}$
production~\cite{CMS:2021rym}. The ATLAS collaboration further
explored the semileptonic reaction $WW/WZ \to \ell \nu q q^\prime$ at
8 TeV~\cite{ATLAS:2017pbb} to extract limits on anomalous TGCs, with
the most stringent constraints arising from cases where a single fat
jet was tagged as a $W$ or $Z$ boson. \smallskip

In this work, we analyze the $WW/WZ$ and $W\gamma$ diboson
productions in the fully hadronic final state, focusing on scenarios where
the boosted $W/Z$ bosons are identified as high transverse momentum
and large-radius jets. We utilize the large-radius jet mass and its
substructure to effectively characterize the hadronically decaying
$W/Z$ bosons and to reduce Standard Model backgrounds. To demonstrate
the potential of these new channels for TGC studies, we recast the
ATLAS searches for heavy resonances decaying into $WW/WZ$ in the
hadronic channel~\cite{ATLAS:2019nat} as well as
$W\gamma$~\cite{ATLAS:2023kcu}. We show that these processes can
lead to TGC bounds comparable to the ones obtained by studying
leptonic final states. Moreover, we also analyze the potential of the
High Luminosity LHC (HL-LHC) run to probe anomalous TGC in  all-hadronic electroweak diboson (EWDB) channels, {\em i.e.} in  the production of pairs $WZ$, $WW$ and $W\gamma$.  \smallskip

This work is organized as follows: in Section~\ref{sec:frame} we
present the adopted theoretical framework as well as the analyses
methodology. Section~\ref{sec:res} contains our results and we
summarize our conclusions in section~\ref{sec:con}.

\section{Analysis Framework}
\label{sec:frame}

Assuming the existence of a mass gap between the new physics energy
scale and the electroweak one, we parametrize the deviations from the
SM TGC predictions using effective field theory. Furthermore, assuming
that the scalar particle observed in 2012~\cite{ATLAS:2012yve,
  CMS:2012qbp} belongs to an electroweak doublet, we can realize the
$SU(2)_L \otimes U(1)_Y$ symmetry linearly, {\em i.e.} we work in the
Standard Model Effective Field Theory (SMEFT) framework. We choose the
Hagiwara, Ishihara, Szalapski, and Zeppenfeld (HISZ) dimension-six
basis~\cite{Hagiwara:1993ck, Hagiwara:1996kf} and we consider three
operators contributing to TGC at dimension-six:
\begin{alignat}{3}
~~~~~~~~~~~~ &\ow =    (D_\mu\Phi)^\dagger\widehat{W}^{\mu\nu}(D_\nu\Phi) \;\;\;\;,\;\;\;\;&&
     \ob =      (D_\mu\Phi)^\dagger\widehat{B}^{\mu\nu}(D_\nu\Phi)
     \;\;\;\;,\;\;\;\; && \owww=  {\rm
       Tr}[\widehat{W}_{\mu}^{\nu}\widehat{W}_{\nu}^{\rho}\widehat{W}_{\rho}^{\mu}]
     \; ,
     \label{eq:tgc}
\end{alignat}
where $\Phi$ stands for the SM Higgs doublet and we have defined
$\widehat{B}_{\mu\nu} \equiv i(g^\prime/2)B_{\mu\nu}$ and
$\widehat{W}_{\mu\nu} \equiv i(g/2)\sigma^aW^a_{\mu\nu}$, with $g$ and
$g^\prime$ being the $SU(2)_L$ and $U(1)_Y$ gauge couplings,
respectively. Here $\sigma^a$ represents the Pauli matrices. In this work we considered the dimension-six effective lagrangian,
\begin{eqnarray}
{\cal L}_{\rm eff} = {\cal L}_{\rm SM}
+ \frac{f_W}{\Lambda^2} {\cal O}_W
+ \frac{f_B}{\Lambda^2} {\cal O}_B
+ \frac{f_{WWW}}{\Lambda^2} {\cal O}_{WWW} \;,
\end{eqnarray}
where $\Lambda$ is the characteristic mass scale of new physics and $f_{W,B,WWW}$ are the Wilson coefficients.\smallskip

The above TGC operators can be qualitatively understood in terms of
the effective $\gamma W^+ W^-$ and $Z W^+W^-$ parametrization
introduced in Ref.~\cite{Hagiwara:1986vm}
\begin{eqnarray}
{\cal L}_{WWV} =
 -i g_{WWV} \Big\{ 
g_1^V \Big( W^+_{\mu\nu} W^{- \, \mu} V^{\nu} 
  - W^+_{\mu} V_{\nu} W^{- \, \mu\nu} \Big)
 + \kappa_V W_\mu^+ W_\nu^- V^{\mu\nu}
+ \frac{\lambda_V}{\widehat M_W^2} W^+_{\mu\nu} W^{- \, \nu\rho} V_\rho^{\; \mu}
 \Big\}
\;\;,
\label{eq:tgc}
\end{eqnarray}
where $V= \gamma, \, Z$, $g_{WW\gamma} = \eh$, $g_{WWZ} = \eh \ch / \sh$, and
$\widehat{M}_W = \eh \vh/2\sh$, with $\eh$ representing the proton
electric charge and $\ch(\sh)$ denoting the sine (cosine) of the weak
mixing angle. In the SM,
$g_1^\gamma=g_1^Z = \kappa_\gamma = \kappa_Z =1$ and
$\lambda_Z=\lambda_\gamma=0$. After including the direct contribution
from the dimension-six operators, electromagnetic gauge invariance
still enforces $g_1^\gamma=1$, while the other effective TGC couplings
read:
\begin{eqnarray}
 &&\Delta  g_1^Z = \frac{\eh^2}{8\sh^2 \ch^2} \frac{\vh^2}{\Lambda^2}
f_W  \;,
  \nonumber
  \\
  && \Delta \kappa_\gamma = \frac{\eh^2}{8\sh^2}
     \frac{\vh^2}{\Lambda^2} \left (  f_W
     +  f_B 
     \right)  \;,
     \nonumber
  \\
  && \Delta \kappa_Z = \frac{\eh^2}{8\sh^2}
     \frac{\vh^2}{\Lambda^2}\left [ f_W - \frac{\sh^2}{\ch^2}
     f_B 
     \right] \;,
     \label{eq:hagi}
  \\
  && \lambda_\gamma = \lambda_Z = \frac{3 \eh^2}{2\sh^2}
     \frac{\widehat M_W^2}{\Lambda^2}
     f_{WWW} \;.
     \nonumber
\end{eqnarray}

In addition to the TGC contributions, diboson production can also be
modified by anomalous couplings of the gauge bosons to
fermions. However, electroweak precision data impose strong
constraints on such couplings~\cite{Butter:2016cvz,
  daSilvaAlmeida:2018iqo, Alves:2018nof, Almeida:2021asy,
  Corbett:2023qtg} and ergo we do not take these contributions
into account. \smallskip

We simulate the EWDB channels at
leading order using
\textsc{MadGraph5\_aMC@NLO}~\cite{Frederix:2018nkq} with the UFO files
for our effective Lagrangian generated with
\textsc{FeynRules}~\cite{Christensen:2008py, Alloul:2013bka}.  We
employ \textsc{PYTHIA8}~\cite{Sjostrand:2007gs} to perform the parton
shower and hadronization, while the fast detector simulation is
carried out with \textsc{Delphes}~\cite{deFavereau:2013fsa}.  Jet
analyses are performed using
\textsc{FASTJET}~\cite{Cacciari:2011ma}. For the EWDB hadronic
channels, the final jets were clustered and trimmed in the same way as
described by the experimental collaborations \cite{ATLAS:2019nat,ATLAS:2023kcu}, using the final state
stable particles after performing the parton-shower and
hadronization. The analysis of the jet-substructure was carried out
using the plugins that are part of the \textsc{FASTJET} contrib
project
(\href{https://fastjet.hepforge.org/contrib/}{https://fastjet.hepforge.org/contrib/}).
\smallskip

\begin{table} [ht]
\begin{tabular}{|@{\hskip 0.5cm}c|l|l|c|l|l|}
\hline
& Channel ($a$) & Distribution & \# bins   &\hspace*{0.2cm} Data set & \hspace*{0.2cm}Int Lum  \\ [0mm]
  \hline
  & $WZ \to \ell^+ \ell^- \ell^{\prime\pm}$ & {$M(WZ)$} & 7& CMS 13
       TeV,  &  137.2 fb$^{-1}$~\cite{CMS:2021lix} \\[0mm]
  \multirow{8}{*}
{\begin{rotate}{90}   EWDB data\end{rotate}}
&$WW \to \ell^+\ell^{(\prime)-}+ 0/1 j$   &$M(\ell^+\ell^{(\prime)-})$
                             &11 & CMS 13 TeV, & 35.9 fb$^{-1}$~\cite{CMS:2020mxy} \\[0mm]
&  $W\gamma \to \ell \nu \gamma$ & $\frac{d^2\sigma}{dp_Td\phi}$ & 12
       & CMS 13 TeV, & 137.1 fb$^{-1}$~\cite{CMS:2021rym}
       \\[0mm]
&  $WW\rightarrow e^\pm \mu^\mp+\sla{E}_T\; (0j)$
&  $m_T$ & 17 (15) &
ATLAS 13 TeV, &36.1 fb$^{-1}$~\cite{Aaboud:2017gsl} \\[0mm]
& $WZ\rightarrow \ell^+\ell^{-}\ell^{(\prime)\pm}$
&  $m_{T}^{WZ}$ & 6
& ATLAS 13 TeV, &36.1 fb$^{-1}$~\cite{ATLAS:2018ogj} \\[0mm]
& $WW\rightarrow \ell^+\ell^{(\prime) -}+\sla{E}_T\; (1j)$ &
$\frac{d\sigma}{dm_{\ell^+\ell^-}}$ & 10 & ATLAS 13 TeV,
 & 139 fb$^{-1}$~\cite{ATLAS:2021jgw} \\[0mm]
  & $WW/WZ \rightarrow 2J$ & $M_{JJ}$ & 11 &
  ATLAS 13 TeV & 139 fb$^{-1}$~\cite{ATLAS:2019nat}
  \\[0mm]
  & $W\gamma \rightarrow J\gamma$ & $M(J\gamma)$ & 20 &
  ATLAS 13 TeV & 139 fb$^{-1}$~\cite{ATLAS:2023kcu} \\[0mm]
    \hline
\end{tabular}
\caption{EWDB data from LHC used in the analyses. For the $W^+ W^-$ results from ATLAS run 2 \cite{Aaboud:2017gsl} we combine the data from the last three bins into one to ensure gaussianity.}
\label{tab:tgv-data}
\end{table}

In this work, we considered two scenarios in our analyses. In the
first scenario, we used the available Run 2 experimental data, which
contains an integrated luminosity of $139$ fb$^{-1}$. In the second
scenario, we performed our analyses for the hadronic EWDB channels
assuming the foreseen integrated luminosity of the High Luminosity LHC
run, {\em i.e.} $3000$ fb$^{-1}$, however, we kept the present
experimental systematic errors. \smallskip

Table~\ref{tab:tgv-data} presents the Run 2 EWDB data used in our
analyses, comprising a total of $92$ data points. In order to improve
the statistical analysis of the hadronic channels, we rebinned the
data. For the $W\gamma \rightarrow J \gamma$ channel, we merged the
bins of the distribution in such a way that the total number of events
in each bin can be described by a Gaussian distribution. For the ATLAS
$WW/WZ \rightarrow 2J$ channel, the bins were combined to ensure at
least one event per bin. In summary, the binnings used in our Run 2
analyses are as follows:\footnote{We maintained the invariant mass
  used in the experimental analyses.}
\begin{eqnarray*}
 && \text{ATLAS } W\gamma\rightarrow J\gamma: \;\; \\
 && (0.80,0.88,0.96,1.04,1.12,1.20,1.28,1.36,1.44,1.52,\\
 &&1.60,1.68,1.76,1.84,1.92,2.00,2.12,2.28,2.56,2.96,7.00)\;\hbox{ TeV} \\
 \\
 && \text{ATLAS } WW/WZ \rightarrow 2J: \;\; \\
 &&(1.30,1.40,1.50,1.60,1.70,1.80,1.90,2.00,2.10,2.40,3.10,5.10) \;\hbox{ TeV}.
\end{eqnarray*}

On the other hand, our choice of bins for the HL-LHC analyses is 
\begin{eqnarray*}
 && \text{ATLAS } W\gamma\rightarrow J\gamma: \;\; \\
 && (0.80, 0.92, 1.00, 1.08, 1.16, 1.24, 1.36, 1.48, 1.60, 1.72, 1.84, 1.96,\\
 &&2.08, 2.20, 2.32, 2.44, 2.56, 2.68, 2.80, 2.96, 3.12, 3.28, 3.48, 3.76, 4.24, 7.00)\;\hbox{ TeV} \\
 \\
 && \text{ATLAS } WW/WZ \rightarrow 2J: \;\; \\
 &&(1.30,1.40,1.60,1.80,2.00,2.20,2.40,2.60, 2.90,3.40,5.10) \;\hbox{ TeV}.
\end{eqnarray*}

\begin{figure}[ht]
        \centering
		\includegraphics[width=0.8\textwidth]{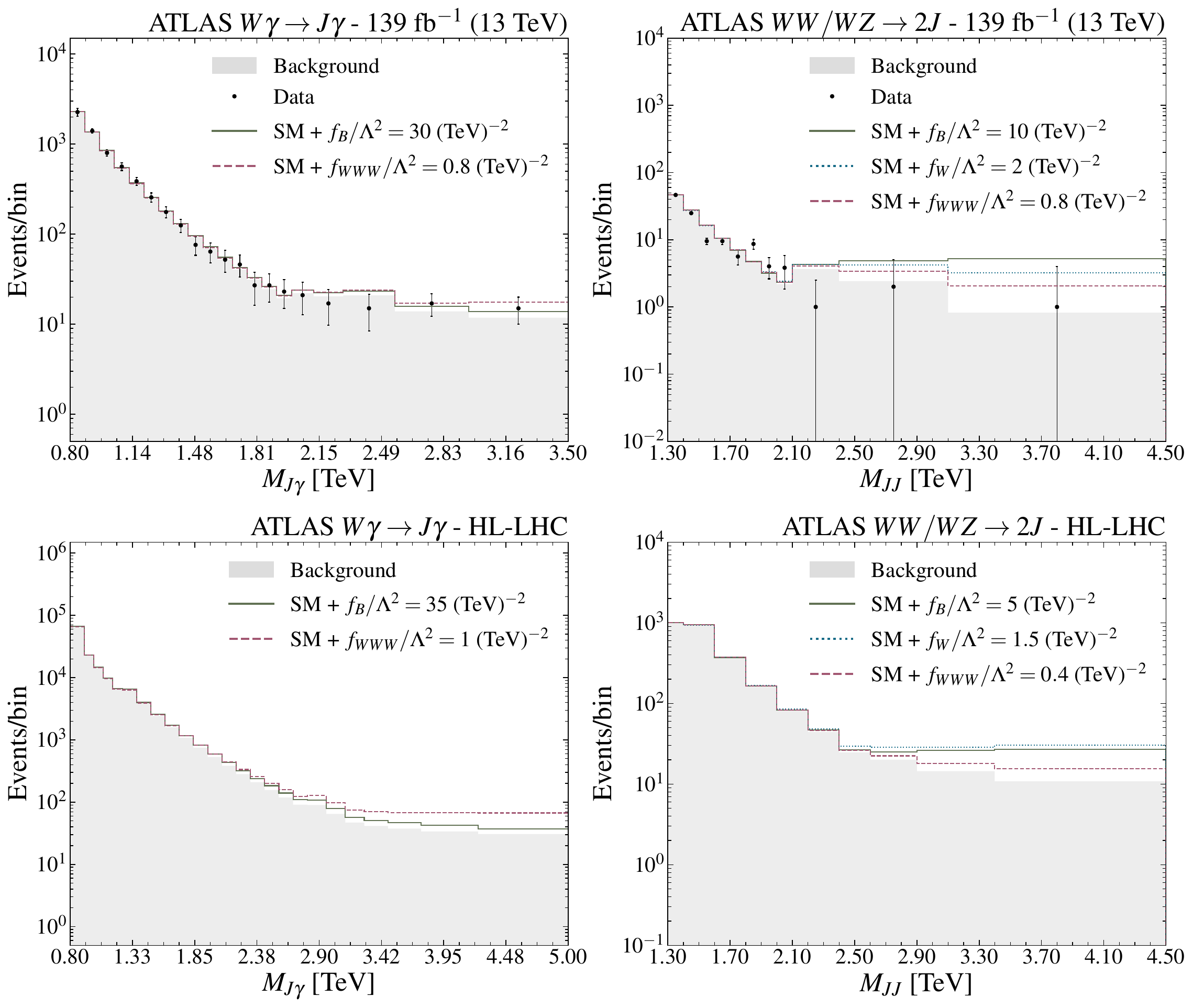}
                \caption{Kinematical distributions employed in the
                  hadronic analyses. The left (right) pannels display
                  the anomalous TGC and background $M_{J\gamma}$
                  ($M_{JJ}$) invariant mass distribution for
                  illustrative values of the dimension-six Wilson
                  coefficients. On the first row, we present the
                  distributions used for the Run 2 analyses, while on
                  the bottom the ones used for the HL-LHC analyses.
                  We used the fit to the background performed in the
                  experimental analyses~\cite{ATLAS:2019nat,
                    ATLAS:2023kcu} for all of our studies, except that
                  in the HL-LHC case, we rescaled it with the proper
                  luminosity. On the left pannels, the last bin
                  contains events up to 7 TeV, while on the right
                  pannels, the last bin contains events up to 5.1
                  TeV.}
        \label{fig:ewdb}
\end{figure}

In Fig.~\ref{fig:ewdb}, we exhibit the anomalous TGC and background
invariant mass distributions for the all hadronic EWDB channels, using
illustrative values of the relevant Wilson coefficients, along with
the experimental data extracted from \cite{ATLAS:2023kcu,
  ATLAS:2019nat} for the Run 2 analysis.  Notice that ${\cal O}_B$ and
${\cal O}_W$ contribute in the same way to $W\gamma$ production; see
Eq.~(\ref{eq:hagi}). As we can see, the presence of anomalous TGC
enhances the cross section at large invariant masses. This behavior is
expected, as the additional TGC contributions spoil the SM high-energy
cancellations. \smallskip

We performed the statistical analysis of the leptonic EWDB data using
the binned chi-squared function described in Ref.~\cite{Corbett:2023qtg}.
On the other hand, for the analysis of the ATLAS $W\gamma\rightarrow J \gamma$ channel,
we profited from the gaussianity of the combined bins and defined the
chi-square function
\begin{eqnarray}
  \chi^2(f_B, f_W, f_{WWW}) = \min_{\Vec{\xi}}\Bigg\{\sum_{i}\Bigg(\dfrac{N_i^\text{obs} - N_i^\text{th}}{\sigma_i}\Bigg)^2 + \xi_1^2 + \xi_2^2\Bigg\},
        \label{eq:chi_square_wgamma}
\end{eqnarray}
where $N_i^\text{obs}$ represents the observed number of events in the
bin $i$ while the theory prediction is given by
\begin{eqnarray}
  N_i^\text{th} = \big(1 + \sigma_{\xi_1} \xi_1\big) N^\text{signal}_i + \big(1 + \sigma_{\xi_2} \xi_2\big) N_i^\text{backg} & \text{with} & N^\text{signal}_i = N^\text{int}_i + N^\text{BSM}_i,
\end{eqnarray}
where $N^\text{backg}_i$ denotes the number of background events
extracted from Ref.~\cite{ATLAS:2023kcu}, $N^\text{int}_i$ stands for
the expected number of events originating from the interference
between the dimension-six operator and SM contributions, and
$N^\text{BSM}_i$ is the pure anomalous contribution to the number of
expected events.  Moreover, the $\sigma_i$ contains the statistical
and background uncertainties added in quadrature, given by
$\sigma_i^2 = N^\text{obs}_i + \sigma_{\text{backg}, i}^2$ where the
last term in this expression was extracted from
Ref.~\cite{ATLAS:2023kcu}. In order to account for possible systematic
theoretical and experimental uncertainties, we defined two
pulls~\cite{Fogli:2002pt}, $\xi_1$ and $\xi_2$, affecting the
normalization of the signal and background, respectively.  The values
chosen for $\sigma_{\xi_1}$ and $\sigma_{\xi_2}$ are 0.2 and 0.3,
respectively. The values for $\sigma_{\xi_1}$ and $\sigma_{\xi_2}$ were estimated analysing the systematic 
errors from Ref. \cite{ATLAS:2023kcu}. Although it is not possible to extract precise uncertainties from the available data, we made conservative estimates. Moreover, we expect the uncertainties on the signal to be slightly smaller than those on the background. In this case, an analytical expression for the pulls can be found by minimizing Eq. (\ref{eq:chi_square_wgamma}) with
respect to $\xi_1$ and $\xi_2$. \smallskip

The statistical analyses of the ATLAS $WW/WZ\rightarrow 2J$ channel
were based on the chi-square function~\cite{Fogli:2002pt}
\begin{eqnarray}
  \chi^2(f_B, f_W, f_{WWW}) = \min_{\Vec{\xi}} \Bigg\{2\sum_i
  \Big[N_i^\text{th} - N_i^\text{obs} + N_i^\text{obs}
  \log\Big(\dfrac{N_i^\text{obs}}{N_i^\text{th}}\Big)\Big] + \xi_1^2 + \xi_2^2 \Bigg\},
\end{eqnarray}
with $N_i^\text{obs}$ standing for the observed number of events in
the $i^\text{th}$ bin and $N_i^\text{th}$ defined as
\begin{eqnarray}
        N_i^\text{th} = \big(1 + \sigma^{\xi_1}_i \xi_1\big) N^\text{signal}_i + 
        \big(1 + \sigma^{\xi_2}_i \xi_2\big) N_i^\text{backg} & 
        \text{with} & N^\text{signal}_i = N^\text{int}_i + N^\text{BSM}_i,
\end{eqnarray}
where $N_i^\text{backg}$, $N_i^\text{int}$ and $N_i^\text{BSM}$
represent the background-fit extracted from Ref. \cite{ATLAS:2019nat},
and the linear and quadratic contributions of the dimension-six
operators, respectively.  The systematic uncertainties were
parameterized by the nuisance parameters $\xi_1$ and $\xi_2$, which
modify the normalization of the signal and background,
respectively. We chose the values for $\sigma^{\xi_2}_i$ to best
represent the experimental errors, and for $\sigma^{\xi_1}_i$, the
values chosen stem from the theoretical uncertainties.  Their values
are
\begin{eqnarray}
  \sigma_i^{\xi_1} &&= (0.05, 0.05, 0.05, 0.05, 0.1, 0.1, 0.1, 0.15,
                      0.15, 0.15, 0.15) \;,\\
  \sigma_i^{\xi_2} &&= (0.1, 0.1, 0.2, 0.2, 0.4, 0.4, 0.5, 0.5, 1.0,
                      1.0, 1.0) \;.
\end{eqnarray}


The HL-LHC analysis is carried out to estimate how the limits on the
Wilson coefficients $f_B$, $f_W$ and $f_{WWW}$, extracted using the
hadronic EWDB channels, can improve with the upcoming LHC runs. Since
there is no available data, we use the SM background fit scaled by a
factor of 3000/139 as the observed number of events. As shown in the
bottom row of Fig. \ref{fig:ewdb}, the number of events for each bin
of the distributions for the $W\gamma$ and $WW/WZ$ channels is
sufficiently large, allowing us to assume gaussianity. To extract the
95\% CL intervals, we use the same statistics defined in
Eq.~(\ref{eq:chi_square_wgamma}), with minor modifications to the
uncertainties. For the ATLAS $W\gamma\rightarrow J\gamma$, we rescaled
the $\sigma_{\text{backg}, i}$ to maintain the current
$\sigma_{\text{backg}, i}/N_i^\text{obs}$ ratio.  For the ATLAS
$WW/WZ\rightarrow 2J$, we defined
$\sigma_{\text{backg}, i} = x_i N^\text{backg}_{i}$, with
\begin{eqnarray}
    x_i = (0.1, 0.1, 0.2, 0.2, 0.4, 0.4, 0.5, 0.5, 1.0, 1.0),
\end{eqnarray}
to represent the same large uncertainties in the background as the
current data.

\section{Results}
\label{sec:res}


To estimate the LHC potential for studying TGC using highly boosted
$W$'s and $Z$'s decaying hadronically, we recast the available ATLAS
data on searches for resonances decaying into $WW/WZ$ and $W\gamma$
pairs followed by the hadronic decays of the gauge
bosons~\cite{ATLAS:2019nat, ATLAS:2023kcu}.  For comparison, we also
obtained the Run 2 limits from combining fully leptonic modes; for
details see Ref.~\cite{Corbett:2023qtg}. \smallskip

Table~\ref{tab:run2-1} presents the 95\% CL allowed intervals for the
three TGC Wilson coefficients contributing to diboson production. In
the second and third columns, we exhibit the results for the analysis
performed using the hadronic $WW/WZ$ production data and the hadronic
$W\gamma$ results, respectively. For the sake of comparison, the
fourth and fifth columns contain the allowed intervals using the
leptonic final states of the $WW/WZ$ and $W\gamma$ diboson
productions. Taking into account only the fully hadronic final states,
the Wilson coefficients $f_B$ and $f_W$ are better constrained by the
$WW/WZ$ diboson production. Additionally, $f_{WWW}$ is more tightly
constrained than $f_W$ and $f_B$, with the $W\gamma$ production
channel leading to the strongest bound. Comparing these results with
those obtained from the leptonic decay modes, we can see that the
fully hadronic channels lead to tighter limits on $f_B$, due to the
$2J$ final state, while a similar limit is obtained for
$f_W$. Moreover, the constraints for $f_{WWW}$ are similar for all
channels. \smallskip

\begin{table}[h]
    \centering
    \renewcommand{\arraystretch}{1.5}
    \begin{tabular}{|c|c|c|c|c|}
      \hline
      \multirow{2}{*}{Coefficient} & \multicolumn{2}{c|}{Hadronic EWDB $(\mathcal{L}=139\rm\, fb^{-1})$} &     \multicolumn{2}{c|}{Leptonic EWDB $(\mathcal{L}=139 \rm\, fb^{-1})$} \\ \cline{2-5}
                                   & ATLAS $WW/WZ \rightarrow 2J$ & ATLAS $W\gamma \rightarrow J\gamma$ & Combined $WW/WZ$ & CMS $W\gamma\rightarrow \ell \nu \gamma$\\
      \hline
      $\dfrac{f_B}{\Lambda^2}$  &  [-8.4, 8.8] & [-33, 33]  & [-13, 15] &  [-17, 19] \\
      $\dfrac{f_W}{\Lambda^2}$  &  [-2.0, 2.3] & [-33, 33] & [-1.3, 2.5] & [-17, 19] \\
      $\dfrac{f_{WWW}}{\Lambda^2}$&[-1.3, 1.3] & [-0.82, 0.82]& [-1.6, 1.6] & [-0.84, 0.74]   \\
      \hline
    \end{tabular}
    \caption{95\% CL intervals for the TGC Wilson coefficients
      originating from the different datasets. We marginalized the
      chi-square with respect the other parameters in the analysis
      with the only exception being the $W\gamma$ channels for which
      we only included $f_B$ since its contribution is identical to
      the $f_W$ one, leading to a blind direction. In the fourth
      column we combined all the available data on $WW$ and $WZ$
      production with the gauge bosons decaying to leptons.  }
    \label{tab:run2-1}
\end{table}

\begin{figure}[ht]
\centering
  \includegraphics[width=0.9\linewidth]{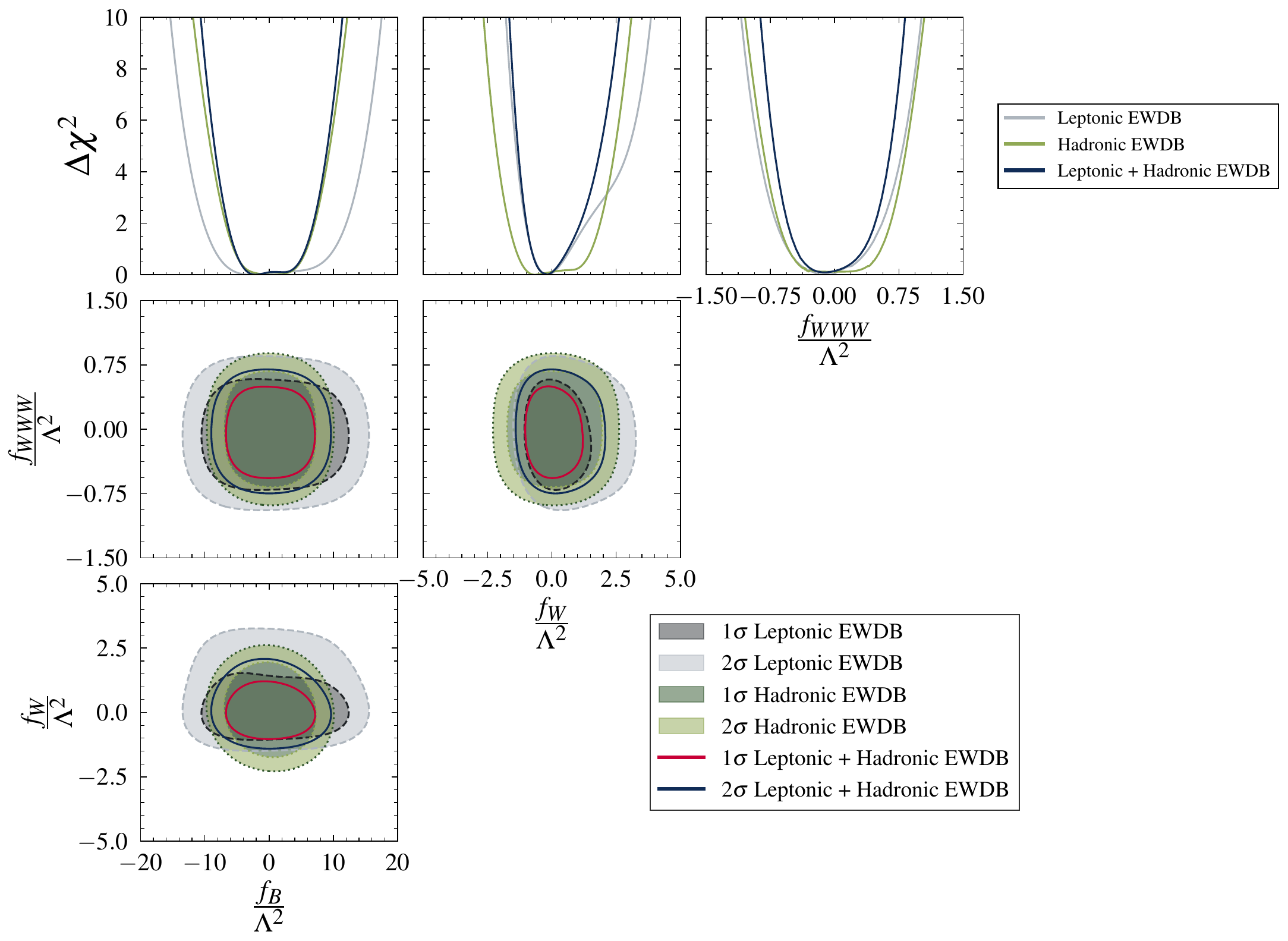}
  \caption {$1\sigma$ and $2\sigma$ allowed regions for each possible pair of 
  Wilson coefficients in the analysis, after marginalizing over 
  the undisplayed parameters, for the different datasets as
  indicated in the figure. We also display the one-dimensional
  marginalized projections of the $\Delta\chi^2$.}
\label{fig:fbfw}
\end{figure}

To compare the impact of the different datasets on the study of
anomalous TGC, Figure~\ref{fig:fbfw} depicts the $1\sigma$ and
$2\sigma$ allowed regions for all parameters in the analyses as well
as the one-dimensional projection of the $\Delta\chi^2$.  As we can
see, the hadronic datasets lead to more stringent constraints on $f_B$
than the leptonic ones, while the bounds on $f_W$ are comparable for
the hadronic and leptonic final states; see Table~\ref{tab:lep-had}
for the marginalized 95\% CL allowed intervals.  For all the hadronic 
(leptonic) results quoted in Table~\ref{tab:lep-had}, we combine all the 
hadronic (leptonic) channels mentioned in Table~\ref{tab:run2-1}. Notably, 
combining the $WW/WZ$ and $W\gamma$ hadronic datasets breaks the blind direction
$f_W + f_B$ that exists in the $W\gamma$ production.  As could have
been anticipated, the combination of the hadronic channels gives
bounds on $f_{WWW}$ similar to the ones originating from the leptonic
final states; see Table~\ref{tab:lep-had}.  The combination of
leptonic and hadronic channels leads, obviously, to more stringent
limits. \smallskip

\begin{table}[h]
    \centering
    \renewcommand{\arraystretch}{1.5}
    \begin{tabular}{|c|c|c|c|}
    \hline
    Coefficients & Hadronic EWDB $(\mathcal{L}=139\rm\, fb^{-1})$ & Leptonic EWDB $(\mathcal{L}=139\rm\, fb^{-1})$ & Leptonic $+$ Hadronic EWDB $(\mathcal{L}=139\rm\, fb^{-1})$ \\
        \hline
         $\dfrac{f_B}{\Lambda^2}$  &  [-8.3, 8.7] & [-12, 12] & [-7.8, 8.4]  \\
         $\dfrac{f_W}{\Lambda^2}$  &  [-2.0, 2.3] & [-1.3, 2.6] & [-1.2, 1.7]\\
         $\dfrac{f_{WWW}}{\Lambda^2}$&[-0.78, 0.78] & [-0.83, 0.72] & [-0.66, 0.60]\\\hline
    \end{tabular}
    \caption{95\% CL intervals for the TGC Wilson coefficients in the
      analysis of the hadronic channels, the result taking into
      account all leptonic channels, as well as the combination of
      leptonic and hadronic channels. In the hadronic (leptonic) results of this table we combine all the hadronic (leptonic) channels mentioned in Table~\ref{tab:run2-1}.}
    \label{tab:lep-had}
\end{table}


We also obtained the attainable limits on the hadronic channel for the
HL-LHC. We assumed that the observed number of events is the one
predict by the fit to the SM background made by the ATLAS
collaboration~\cite{ATLAS:2019nat, ATLAS:2023kcu}. We also kept the
same systematic uncertainties of this fit for the LHC Run 2 and we
considered an integrated luminosity of 3000 fb$^{-1}$. Our results are
presented in Table~\ref{tab:hllhc}. As we can see, the HL-LHC can
improve the present limits on the $f_B$ and $f_{WWW}$ coefficients
(see last column of Table \ref{tab:lep-had}) by almost a factor of 1.5
when we consider only the hadronic channels in contrast with the
current Leptonic limits. \smallskip

\begin{table}[h]
    \centering
    \renewcommand{\arraystretch}{1.5}
    \begin{tabular}{|c|c|c|c|}
        \hline
        \multirow{2}{*}{Coefficient} & \multicolumn{3}{c|}{Hadronic EWDB ($\mathcal{L}=3000\rm\, fb^{-1})$} \\ \cline{2-4}
        & ATLAS $WW/WZ$ & ATLAS $W\gamma$ & Combined $WW/WZ$ and $W\gamma$ \\ \hline
        $\dfrac{f_B}{\Lambda^2}$  &  [-5.3, 5.6] & [-20, 20] & [-5.3, 5.6]\\
        $\dfrac{f_W}{\Lambda^2}$  &  [-1.2, 1.5] & [-20, 20] & [-1.2, 1.5]\\
        $\dfrac{f_{WWW}}{\Lambda^2}$  &  [-0.81, 0.81] & [-0.40, 0.40] & [--0.40, 0.40]\\\hline
    \end{tabular}
    \caption{95\% CL allowed intervals for the TGC Wilson coefficients
      for the hadronic model $WW/WZ$ and $W\gamma$ as well as their
      combination. We assumed an integrated luminosity of 3000
      fb$^{-1}$.}
    \label{tab:hllhc}
\end{table}

\section{Final remarks}
\label{sec:con}

We analyzed the LHC potential to probe anomalous TGC using the $WW/WZ$
and $W\gamma$ channels when the $W$'s and $Z$'s decay hadronically.
We considered boosted final states where the $W/Z$ decay products give
rise to a fat jet. In our analyses, we performed the same sequence of
cuts used by the ATLAS collaboration for the search of heavy
resonances~\cite{ATLAS:2019nat, ATLAS:2023kcu}. We also considered the
SM background as evaluated in the experimental studies. \smallskip

To gauge our results, we compared the limits of the TGC
Wilson coefficients derived from the fully hadronic mode with the ones
from the leptonic final state. Our results indicate that the limits from leptonic and hadronic channels are similar. Ergo, the
addition of the hadronic mode to the TGC analysis will lead to more
stringent global fits. \smallskip


Finally, we should conclude by reiterating the fact that the results presented in this paper relies heavily on the experimental fit to the SM background and the fits have large systematic uncertainties. In fact, this even leaves a room for improvement if the systematic uncertainties can be reduced for the high luminosity run of the LHC.

\acknowledgments
We would like to thank Najimuddin Khan for useful discussions in the early part of the project.
OJPE is partially supported by CNPq grant number 305762/2019-2 and
FAPESP grant 2019/04837-9.  TG would like to acknowledge support from the Department of Atomic Energy, Government of India, for Harish-Chandra Research Institute. MM is supported 
by FAPESP grant number 2022/11293-8. SS is supported by FAPESP Grant number 2021/09547-9.

\clearpage
\bibliography{references}

\begin{thebibliography}{29}
\expandafter\ifx\csname natexlab\endcsname\relax\def\natexlab#1{#1}\fi
\expandafter\ifx\csname bibnamefont\endcsname\relax
  \def\bibnamefont#1{#1}\fi
\expandafter\ifx\csname bibfnamefont\endcsname\relax
  \def\bibfnamefont#1{#1}\fi
\expandafter\ifx\csname citenamefont\endcsname\relax
  \def\citenamefont#1{#1}\fi
\expandafter\ifx\csname url\endcsname\relax
  \def\url#1{\texttt{#1}}\fi
\expandafter\ifx\csname urlprefix\endcsname\relax\def\urlprefix{URL }\fi
\providecommand{\bibinfo}[2]{#2}
\providecommand{\eprint}[2][]{\url{#2}}

\bibitem[{\citenamefont{Arzt et~al.}(1995)\citenamefont{Arzt, Einhorn, and
  Wudka}}]{Arzt:1994gp}
\bibinfo{author}{\bibfnamefont{C.}~\bibnamefont{Arzt}},
  \bibinfo{author}{\bibfnamefont{M.~B.} \bibnamefont{Einhorn}},
  \bibnamefont{and} \bibinfo{author}{\bibfnamefont{J.}~\bibnamefont{Wudka}},
  \bibinfo{journal}{Nucl. Phys. B} \textbf{\bibinfo{volume}{433}},
  \bibinfo{pages}{41} (\bibinfo{year}{1995}), \eprint{hep-ph/9405214}.

\bibitem[{\citenamefont{Falkowski et~al.}(2017)\citenamefont{Falkowski,
  Gonzalez-Alonso, Greljo, Marzocca, and Son}}]{Falkowski:2016cxu}
\bibinfo{author}{\bibfnamefont{A.}~\bibnamefont{Falkowski}},
  \bibinfo{author}{\bibfnamefont{M.}~\bibnamefont{Gonzalez-Alonso}},
  \bibinfo{author}{\bibfnamefont{A.}~\bibnamefont{Greljo}},
  \bibinfo{author}{\bibfnamefont{D.}~\bibnamefont{Marzocca}}, \bibnamefont{and}
  \bibinfo{author}{\bibfnamefont{M.}~\bibnamefont{Son}},
  \bibinfo{journal}{JHEP} \textbf{\bibinfo{volume}{02}}, \bibinfo{pages}{115}
  (\bibinfo{year}{2017}), \eprint{1609.06312}.

\bibitem[{\citenamefont{Arkani-Hamed et~al.}(2001)\citenamefont{Arkani-Hamed,
  Cohen, and Georgi}}]{Arkani-Hamed:2001kyx}
\bibinfo{author}{\bibfnamefont{N.}~\bibnamefont{Arkani-Hamed}},
  \bibinfo{author}{\bibfnamefont{A.~G.} \bibnamefont{Cohen}}, \bibnamefont{and}
  \bibinfo{author}{\bibfnamefont{H.}~\bibnamefont{Georgi}},
  \bibinfo{journal}{Phys. Rev. Lett.} \textbf{\bibinfo{volume}{86}},
  \bibinfo{pages}{4757} (\bibinfo{year}{2001}), \eprint{hep-th/0104005}.

\bibitem[{\citenamefont{{{ATLAS Collaboration}}}(2018)}]{ATLAS:2018ogj}
\bibinfo{author}{\bibnamefont{{{ATLAS Collaboration}}}} (\bibinfo{year}{2018}),
  \bibinfo{note}{{ ATLAS-CONF-2018-034 ,
  \url{https://cds.cern.ch/record/2630187}}}.

\bibitem[{\citenamefont{{{CMS
  Collaboration}}}(2021{\natexlab{a}})}]{CMS:2021lix}
\bibinfo{author}{\bibnamefont{{{CMS Collaboration}}}}
  (\bibinfo{year}{2021}{\natexlab{a}}), \bibinfo{note}{{CMS-PAS-SMP-20-014,
  \url{https://cds.cern.ch/record/2758362}}}.

\bibitem[{\citenamefont{Sirunyan et~al.}(2020)}]{CMS:2020mxy}
\bibinfo{author}{\bibfnamefont{A.~M.} \bibnamefont{Sirunyan}}
  \bibnamefont{et~al.} (\bibinfo{collaboration}{CMS}), \bibinfo{journal}{Phys.
  Rev. D} \textbf{\bibinfo{volume}{102}}, \bibinfo{pages}{092001}
  (\bibinfo{year}{2020}), \eprint{2009.00119}.

\bibitem[{\citenamefont{Aad et~al.}(2021)}]{ATLAS:2021jgw}
\bibinfo{author}{\bibfnamefont{G.}~\bibnamefont{Aad}} \bibnamefont{et~al.}
  (\bibinfo{collaboration}{ATLAS}), \bibinfo{journal}{JHEP}
  \textbf{\bibinfo{volume}{06}}, \bibinfo{pages}{003} (\bibinfo{year}{2021}),
  \eprint{2103.10319}.

\bibitem[{\citenamefont{{{CMS
  Collaboration}}}(2021{\natexlab{b}})}]{CMS:2021rym}
\bibinfo{author}{\bibnamefont{{{CMS Collaboration}}}}
  (\bibinfo{year}{2021}{\natexlab{b}}), \bibinfo{note}{{CMS-PAS-SMP-20-005,
  \url{https://cds.cern.ch/record/2757267}}}.

\bibitem[{\citenamefont{Aaboud et~al.}(2017)}]{ATLAS:2017pbb}
\bibinfo{author}{\bibfnamefont{M.}~\bibnamefont{Aaboud}} \bibnamefont{et~al.}
  (\bibinfo{collaboration}{ATLAS}), \bibinfo{journal}{Eur. Phys. J. C}
  \textbf{\bibinfo{volume}{77}}, \bibinfo{pages}{563} (\bibinfo{year}{2017}),
  \eprint{1706.01702}.

\bibitem[{\citenamefont{Aad et~al.}(2019)}]{ATLAS:2019nat}
\bibinfo{author}{\bibfnamefont{G.}~\bibnamefont{Aad}} \bibnamefont{et~al.}
  (\bibinfo{collaboration}{ATLAS}), \bibinfo{journal}{JHEP}
  \textbf{\bibinfo{volume}{09}}, \bibinfo{pages}{091} (\bibinfo{year}{2019}),
  \bibinfo{note}{[Erratum: JHEP 06, 042 (2020)]}, \eprint{1906.08589}.

\bibitem[{\citenamefont{Aad et~al.}(2023)}]{ATLAS:2023kcu}
\bibinfo{author}{\bibfnamefont{G.}~\bibnamefont{Aad}} \bibnamefont{et~al.}
  (\bibinfo{collaboration}{ATLAS}), \bibinfo{journal}{JHEP}
  \textbf{\bibinfo{volume}{07}}, \bibinfo{pages}{125} (\bibinfo{year}{2023}),
  \eprint{2304.11962}.

\bibitem[{\citenamefont{Aad et~al.}(2012)}]{ATLAS:2012yve}
\bibinfo{author}{\bibfnamefont{G.}~\bibnamefont{Aad}} \bibnamefont{et~al.}
  (\bibinfo{collaboration}{ATLAS}), \bibinfo{journal}{Phys. Lett. B}
  \textbf{\bibinfo{volume}{716}}, \bibinfo{pages}{1} (\bibinfo{year}{2012}),
  \eprint{1207.7214}.

\bibitem[{\citenamefont{Chatrchyan et~al.}(2012)}]{CMS:2012qbp}
\bibinfo{author}{\bibfnamefont{S.}~\bibnamefont{Chatrchyan}}
  \bibnamefont{et~al.} (\bibinfo{collaboration}{CMS}), \bibinfo{journal}{Phys.
  Lett. B} \textbf{\bibinfo{volume}{716}}, \bibinfo{pages}{30}
  (\bibinfo{year}{2012}), \eprint{1207.7235}.

\bibitem[{\citenamefont{Hagiwara et~al.}(1993)\citenamefont{Hagiwara, Ishihara,
  Szalapski, and Zeppenfeld}}]{Hagiwara:1993ck}
\bibinfo{author}{\bibfnamefont{K.}~\bibnamefont{Hagiwara}},
  \bibinfo{author}{\bibfnamefont{S.}~\bibnamefont{Ishihara}},
  \bibinfo{author}{\bibfnamefont{R.}~\bibnamefont{Szalapski}},
  \bibnamefont{and}
  \bibinfo{author}{\bibfnamefont{D.}~\bibnamefont{Zeppenfeld}},
  \bibinfo{journal}{Phys. Rev.} \textbf{\bibinfo{volume}{D48}},
  \bibinfo{pages}{2182} (\bibinfo{year}{1993}).

\bibitem[{\citenamefont{Hagiwara et~al.}(1997)\citenamefont{Hagiwara,
  Hatsukano, Ishihara, and Szalapski}}]{Hagiwara:1996kf}
\bibinfo{author}{\bibfnamefont{K.}~\bibnamefont{Hagiwara}},
  \bibinfo{author}{\bibfnamefont{T.}~\bibnamefont{Hatsukano}},
  \bibinfo{author}{\bibfnamefont{S.}~\bibnamefont{Ishihara}}, \bibnamefont{and}
  \bibinfo{author}{\bibfnamefont{R.}~\bibnamefont{Szalapski}},
  \bibinfo{journal}{Nucl. Phys.} \textbf{\bibinfo{volume}{B496}},
  \bibinfo{pages}{66} (\bibinfo{year}{1997}), \eprint{hep-ph/9612268}.

\bibitem[{\citenamefont{Hagiwara et~al.}(1987)\citenamefont{Hagiwara, Peccei,
  Zeppenfeld, and Hikasa}}]{Hagiwara:1986vm}
\bibinfo{author}{\bibfnamefont{K.}~\bibnamefont{Hagiwara}},
  \bibinfo{author}{\bibfnamefont{R.~D.} \bibnamefont{Peccei}},
  \bibinfo{author}{\bibfnamefont{D.}~\bibnamefont{Zeppenfeld}},
  \bibnamefont{and} \bibinfo{author}{\bibfnamefont{K.}~\bibnamefont{Hikasa}},
  \bibinfo{journal}{Nucl. Phys.} \textbf{\bibinfo{volume}{B282}},
  \bibinfo{pages}{253} (\bibinfo{year}{1987}).

\bibitem[{\citenamefont{Butter et~al.}(2016)\citenamefont{Butter, \'Eboli,
  Gonzalez-Fraile, Gonzalez-Garcia, Plehn, and Rauch}}]{Butter:2016cvz}
\bibinfo{author}{\bibfnamefont{A.}~\bibnamefont{Butter}},
  \bibinfo{author}{\bibfnamefont{O.~J.~P.} \bibnamefont{\'Eboli}},
  \bibinfo{author}{\bibfnamefont{J.}~\bibnamefont{Gonzalez-Fraile}},
  \bibinfo{author}{\bibfnamefont{M.~C.} \bibnamefont{Gonzalez-Garcia}},
  \bibinfo{author}{\bibfnamefont{T.}~\bibnamefont{Plehn}}, \bibnamefont{and}
  \bibinfo{author}{\bibfnamefont{M.}~\bibnamefont{Rauch}},
  \bibinfo{journal}{JHEP} \textbf{\bibinfo{volume}{07}}, \bibinfo{pages}{152}
  (\bibinfo{year}{2016}), \eprint{1604.03105}.

\bibitem[{\citenamefont{da~Silva~Almeida
  et~al.}(2019)\citenamefont{da~Silva~Almeida, Alves, Rosa~Agostinho, \'Eboli,
  and Gonzalez-Garcia}}]{daSilvaAlmeida:2018iqo}
\bibinfo{author}{\bibfnamefont{E.}~\bibnamefont{da~Silva~Almeida}},
  \bibinfo{author}{\bibfnamefont{A.}~\bibnamefont{Alves}},
  \bibinfo{author}{\bibfnamefont{N.}~\bibnamefont{Rosa~Agostinho}},
  \bibinfo{author}{\bibfnamefont{O.~J.~P.} \bibnamefont{\'Eboli}},
  \bibnamefont{and} \bibinfo{author}{\bibfnamefont{M.~C.}
  \bibnamefont{Gonzalez-Garcia}}, \bibinfo{journal}{Phys. Rev. D}
  \textbf{\bibinfo{volume}{99}}, \bibinfo{pages}{033001}
  (\bibinfo{year}{2019}), \eprint{1812.01009}.

\bibitem[{\citenamefont{Alves et~al.}(2018)\citenamefont{Alves, Rosa-Agostinho,
  \'Eboli, and Gonzalez-Garcia}}]{Alves:2018nof}
\bibinfo{author}{\bibfnamefont{A.}~\bibnamefont{Alves}},
  \bibinfo{author}{\bibfnamefont{N.}~\bibnamefont{Rosa-Agostinho}},
  \bibinfo{author}{\bibfnamefont{O.~J.~P.} \bibnamefont{\'Eboli}},
  \bibnamefont{and} \bibinfo{author}{\bibfnamefont{M.~C.}
  \bibnamefont{Gonzalez-Garcia}}, \bibinfo{journal}{Phys. Rev. D}
  \textbf{\bibinfo{volume}{98}}, \bibinfo{pages}{013006}
  (\bibinfo{year}{2018}), \eprint{1805.11108}.

\bibitem[{\citenamefont{Almeida et~al.}(2021)\citenamefont{Almeida, Alves,
  \'Eboli, and Gonzalez-Garcia}}]{Almeida:2021asy}
\bibinfo{author}{\bibfnamefont{E.~d.~S.} \bibnamefont{Almeida}},
  \bibinfo{author}{\bibfnamefont{A.}~\bibnamefont{Alves}},
  \bibinfo{author}{\bibfnamefont{O.~J.~P.} \bibnamefont{\'Eboli}},
  \bibnamefont{and} \bibinfo{author}{\bibfnamefont{M.~C.}
  \bibnamefont{Gonzalez-Garcia}} (\bibinfo{year}{2021}), \eprint{2108.04828}.

\bibitem[{\citenamefont{Corbett et~al.}(2023)\citenamefont{Corbett, Desai,
  \'Eboli, Gonzalez-Garcia, Martines, and Reimitz}}]{Corbett:2023qtg}
\bibinfo{author}{\bibfnamefont{T.}~\bibnamefont{Corbett}},
  \bibinfo{author}{\bibfnamefont{J.}~\bibnamefont{Desai}},
  \bibinfo{author}{\bibfnamefont{O.~J.~P.} \bibnamefont{\'Eboli}},
  \bibinfo{author}{\bibfnamefont{M.~C.} \bibnamefont{Gonzalez-Garcia}},
  \bibinfo{author}{\bibfnamefont{M.}~\bibnamefont{Martines}}, \bibnamefont{and}
  \bibinfo{author}{\bibfnamefont{P.}~\bibnamefont{Reimitz}},
  \bibinfo{journal}{Phys. Rev. D} \textbf{\bibinfo{volume}{107}},
  \bibinfo{pages}{115013} (\bibinfo{year}{2023}), \eprint{2304.03305}.

\bibitem[{\citenamefont{Frederix et~al.}(2018)\citenamefont{Frederix, Frixione,
  Hirschi, Pagani, Shao, and Zaro}}]{Frederix:2018nkq}
\bibinfo{author}{\bibfnamefont{R.}~\bibnamefont{Frederix}},
  \bibinfo{author}{\bibfnamefont{S.}~\bibnamefont{Frixione}},
  \bibinfo{author}{\bibfnamefont{V.}~\bibnamefont{Hirschi}},
  \bibinfo{author}{\bibfnamefont{D.}~\bibnamefont{Pagani}},
  \bibinfo{author}{\bibfnamefont{H.~S.} \bibnamefont{Shao}}, \bibnamefont{and}
  \bibinfo{author}{\bibfnamefont{M.}~\bibnamefont{Zaro}},
  \bibinfo{journal}{JHEP} \textbf{\bibinfo{volume}{07}}, \bibinfo{pages}{185}
  (\bibinfo{year}{2018}), \eprint{1804.10017}.

\bibitem[{\citenamefont{Christensen and Duhr}(2009)}]{Christensen:2008py}
\bibinfo{author}{\bibfnamefont{N.~D.} \bibnamefont{Christensen}}
  \bibnamefont{and} \bibinfo{author}{\bibfnamefont{C.}~\bibnamefont{Duhr}},
  \bibinfo{journal}{Comput. Phys. Commun.} \textbf{\bibinfo{volume}{180}},
  \bibinfo{pages}{1614} (\bibinfo{year}{2009}), \eprint{0806.4194}.

\bibitem[{\citenamefont{Alloul et~al.}(2014)\citenamefont{Alloul, Christensen,
  Degrande, Duhr, and Fuks}}]{Alloul:2013bka}
\bibinfo{author}{\bibfnamefont{A.}~\bibnamefont{Alloul}},
  \bibinfo{author}{\bibfnamefont{N.~D.} \bibnamefont{Christensen}},
  \bibinfo{author}{\bibfnamefont{C.}~\bibnamefont{Degrande}},
  \bibinfo{author}{\bibfnamefont{C.}~\bibnamefont{Duhr}}, \bibnamefont{and}
  \bibinfo{author}{\bibfnamefont{B.}~\bibnamefont{Fuks}},
  \bibinfo{journal}{Comput. Phys. Commun.} \textbf{\bibinfo{volume}{185}},
  \bibinfo{pages}{2250} (\bibinfo{year}{2014}), \eprint{1310.1921}.

\bibitem[{\citenamefont{Sjostrand et~al.}(2008)\citenamefont{Sjostrand, Mrenna,
  and Skands}}]{Sjostrand:2007gs}
\bibinfo{author}{\bibfnamefont{T.}~\bibnamefont{Sjostrand}},
  \bibinfo{author}{\bibfnamefont{S.}~\bibnamefont{Mrenna}}, \bibnamefont{and}
  \bibinfo{author}{\bibfnamefont{P.~Z.} \bibnamefont{Skands}},
  \bibinfo{journal}{Comput. Phys. Commun.} \textbf{\bibinfo{volume}{178}},
  \bibinfo{pages}{852} (\bibinfo{year}{2008}), \eprint{0710.3820}.

\bibitem[{\citenamefont{de~Favereau et~al.}(2014)\citenamefont{de~Favereau,
  Delaere, Demin, Giammanco, Lemaitre, Mertens, and
  Selvaggi}}]{deFavereau:2013fsa}
\bibinfo{author}{\bibfnamefont{J.}~\bibnamefont{de~Favereau}},
  \bibinfo{author}{\bibfnamefont{C.}~\bibnamefont{Delaere}},
  \bibinfo{author}{\bibfnamefont{P.}~\bibnamefont{Demin}},
  \bibinfo{author}{\bibfnamefont{A.}~\bibnamefont{Giammanco}},
  \bibinfo{author}{\bibfnamefont{V.}~\bibnamefont{Lemaitre}},
  \bibinfo{author}{\bibfnamefont{A.}~\bibnamefont{Mertens}}, \bibnamefont{and}
  \bibinfo{author}{\bibfnamefont{M.}~\bibnamefont{Selvaggi}}
  (\bibinfo{collaboration}{DELPHES 3}), \bibinfo{journal}{JHEP}
  \textbf{\bibinfo{volume}{02}}, \bibinfo{pages}{057} (\bibinfo{year}{2014}),
  \eprint{1307.6346}.

\bibitem[{\citenamefont{Cacciari et~al.}(2012)\citenamefont{Cacciari, Salam,
  and Soyez}}]{Cacciari:2011ma}
\bibinfo{author}{\bibfnamefont{M.}~\bibnamefont{Cacciari}},
  \bibinfo{author}{\bibfnamefont{G.~P.} \bibnamefont{Salam}}, \bibnamefont{and}
  \bibinfo{author}{\bibfnamefont{G.}~\bibnamefont{Soyez}},
  \bibinfo{journal}{Eur. Phys. J. C} \textbf{\bibinfo{volume}{72}},
  \bibinfo{pages}{1896} (\bibinfo{year}{2012}), \eprint{1111.6097}.

\bibitem[{\citenamefont{Aaboud et~al.}(2018)}]{Aaboud:2017gsl}
\bibinfo{author}{\bibfnamefont{M.}~\bibnamefont{Aaboud}} \bibnamefont{et~al.}
  (\bibinfo{collaboration}{ATLAS}), \bibinfo{journal}{Eur. Phys. J.}
  \textbf{\bibinfo{volume}{C78}}, \bibinfo{pages}{24} (\bibinfo{year}{2018}),
  \eprint{1710.01123}.

\bibitem[{\citenamefont{Fogli et~al.}(2002)\citenamefont{Fogli, Lisi, Marrone,
  Montanino, and Palazzo}}]{Fogli:2002pt}
\bibinfo{author}{\bibfnamefont{G.~L.} \bibnamefont{Fogli}},
  \bibinfo{author}{\bibfnamefont{E.}~\bibnamefont{Lisi}},
  \bibinfo{author}{\bibfnamefont{A.}~\bibnamefont{Marrone}},
  \bibinfo{author}{\bibfnamefont{D.}~\bibnamefont{Montanino}},
  \bibnamefont{and} \bibinfo{author}{\bibfnamefont{A.}~\bibnamefont{Palazzo}},
  \bibinfo{journal}{Phys. Rev. D} \textbf{\bibinfo{volume}{66}},
  \bibinfo{pages}{053010} (\bibinfo{year}{2002}), \eprint{hep-ph/0206162}.

\end{thebibliography}
\end{document}